\documentclass[preprint]{aastex}

\usepackage{bm}
\usepackage{amsmath, amsthm, amssymb}

\shorttitle{U Scorpii}
\shortauthors{Martin, Livio \& Schaefer}

\begin{document}

\title{On the Orbital Period Change in the Recurrent Nova U Scorpii }

\author{Rebecca G. Martin\altaffilmark{1}, Mario Livio\altaffilmark{1}
  and Bradley E. Schaefer\altaffilmark{2}} 

\altaffiltext{1}{Space Telescope Science Institute, 3700 San Martin
  Drive, Baltimore, MD 21218}
\altaffiltext{2}{Physics and Astronomy, Louisiana State University, Baton Rouge, LA, 70803}

\begin{abstract}
 The orbital period of the recurrent nova U Sco has been observed to
 decrease during the 1999 outburst. In an outburst mass is ejected
 from the surface of the white dwarf. The separation of the binary
 system widens and the orbital period increases. We find that magnetic
 braking between outbursts, mass transfer to the companion, and
 frictional angular momentum losses during outbursts are all too small
 to account for this unexpected change. We find, however, that if the
 secondary has a sufficiently strong magnetic field, $B \approx
 8\times 10^3\,\rm G$, then the ejected material can couple to it
 and corrotate with the system. The ejected material gains angular
 momentum while the binary system loses it and the period
 decreases. If such a strong magnetic field is indeed present, then we
 predict that a period decrease should be observed also during the
 current 2010 outburst. If, however, the presence of such a field can
 be ruled out observationally, then the cause for the period decrease
 (if confirmed) remains unknown.
\end{abstract}

\keywords{stars: binaries - stars: magnetic - stars: novae - stars
  individual (U Scorpii)}

\section{Introduction}

Recurrent novae are cataclysmic variables with outbursts at intervals
of $10-80\,\rm yr$ \citep{warner95, webbink87}. They are binary
systems in which mass is transferred from a main-sequence star or a
red giant to a white dwarf by Roche-lobe overflow. The critical amount
of mass that can be accreted on to the surface of a white dwarf prior
to an outburst is a strongly decreasing function of the white dwarf
mass \citep{truran86}. At this mass limit, the temperature and density
at the base of the accreted layer are high enough for hydrogen to
ignite. The temperature then rises rapidly in a thermonuclear runaway
\citep{starrfield88} and the pressure becomes high enough, so that
aided by radioactive decays, most of the accreted material is ejected.
To account for the short timescale between the outbursts, the white
dwarf in a recurrent nova system must have a mass close to the
Chandrasekhar limit \citep[e.g.][]{kato88,kato89}.

During an outburst a finite amount of material is expelled on a short
timescale of a few months. Because of the mass loss from the system,
the binary separation widens and so the orbital period
increases. Angular momentum is continually lost from the system
between outbursts, mainly because of gravitational radiation for the
closest systems and magnetic braking for the wider systems. These
mechanisms both cause the period to decrease on a very long
timescale. During the outburst itself the period can decrease, in
systems with orbital periods shorter than about 8 hours, because a
small amount of mass can be transferred to the companion
\citep{shara86} and also because of frictional angular momentum losses
as the binary moves through the ejected material \citep{livio91}.

There are now ten recurrent novae known in our galaxy \citep[the tenth
  one was discovered last year;][]{pagnotta09} and one system in the
LMC. In this group, U Sco has the fastest decline rate of the light
curve in past outbursts, and the shortest recurrence period
\citep[$11\,\rm yr$ since the last outburst,][]{schaefer01}. It has
outbursts recorded in 1863, 1906, 1917, 1936, 1945, 1969, 1979, 1987,
1999 \citep{schaefer10a} and 2010 \citep{schaefer10b} and other have
likely been missed because of its proximity to the Sun
\citep{schaefer04}. The companion to the white dwarf in the system is
a subgiant \citep{schaefer90}.

The 1999 outburst was detected by \cite{schmeer99}. In this outburst,
the orbital period of U Sco has been observed to decrease (Schaefer,
unpublished). Since the orbital period of U Sco is about 30 hours, it
was expected to increase during an outburst \citep{livio91} and so in
this paper we investigate how a period decrease could occur.

\section{Outburst Model}

We first consider a simple model of the outburst where the ejected
material carries away the specific angular momentum of the white
dwarf.  We assume that all the the material that has been accreted
since the last outburst is ejected in the outburst and we consider the
change to the orbital period when the mass is ejected.

The non-degenerate mass accumulated on to the surface of the white
dwarf is very thin and so the pressure at the base of the layer is
given approximately by
\begin{equation}
P=\frac{G M_1 \Delta m}{R_1^2}\frac{1}{4\pi R_1^2}
\end{equation}
where $\Delta m$ is the accumulated mass and the white dwarf has mass
$M_1$ and radius $R_1$. The envelope is ejected when the pressure at the
surface of the white dwarf reaches a critical value of the order of
$P_{\rm crit}=10^{20}\,\rm dyn\,cm^{-2}$
\citep[e.g.][]{fujimoto82a,fujimoto82b,macdonald83}. We find that the
amount of mass that accumulates before a nova outburst is of order
\begin{equation}
\Delta m=4\pi R_1^4 \frac{P_{\rm crit}}{G M_1}.
\label{dm}
\end{equation}

The angular momentum of the binary star system is given by
\begin{equation}
J=\frac{M_{2} M_1}{M}a^2 \Omega 
\label{da}
\end{equation}
where $a$ is the separation of the two stars and the angular velocity,
$\Omega$, is given by Kepler's law
\begin{equation}
\Omega^2=\frac{GM}{a^3}.
\label{kepler}
\end{equation}
Here the mass of the companion star is $M_2$ and the total mass of the
system is $M=M_1+M_2$.  Then we can express the angular momentum of
the binary system as
\begin{equation}
J= \frac{M_{2} M_1}{M^\frac{1}{3}}G^\frac{2}{3}\left(\frac{P}{2\pi}\right)^\frac{1}{3}
=\frac{M_{2}M_1}{M^\frac{1}{2}}G^\frac{1}{2}a^\frac{1}{2}.
\label{angmom}
\end{equation}
With this we can find the change in the angular momentum of the system
for a given period and a given change in the mass.

If the mass carries away its specific angular momentum then the
angular momentum loss from the system is
\begin{equation}
\Delta J=-\Delta m \,a_1^2 \Omega
\label{dj}
\end{equation}
where $a_1$ is the distance of $M_1$ to the center of mass of the binary
\begin{equation}
a_1=\frac{M_{2}}{M}a
\end{equation}
so that with equations~(\ref{da}) and~(\ref{dj}) we find
\begin{equation}
\frac{\Delta J}{J}=-\frac{\Delta m}{M}\frac{M_2}{M_1}.
\label{loss}
\end{equation}
By differentiating equation~(\ref{angmom}) we also have
\begin{equation}
\frac{\Delta J}{J}=\frac{\Delta M_1}{M_1}-\frac{1}{2}\frac{\Delta M}{M}+\frac{1}{2}\frac{\Delta a}{a}
\label{change}
\end{equation}
where $\Delta M_1=\Delta M=-\Delta m$ and $\Delta a$ is the
corresponding change in the separation of the system (due to mass
lost) during the outburst.  Equating~(\ref{loss})
and~(\ref{change}) we find
\begin{equation}
\frac{\Delta a}{a}=\frac{\Delta m}{M}.
\label{over}
\end{equation}
As mass is lost from the system in the outburst the separation increases.

By differentiating equation~(\ref{kepler}) we find the period change
during the outburst to be
\begin{equation}
\frac{\Delta P}{P}=-\frac{\Delta \Omega}{\Omega}
=-\frac{1}{2}\frac{\Delta M}{M}+\frac{3}{2}\frac{\Delta a}{a}
\end{equation}
and with equation~(\ref{over}) we find
\begin{equation}
\frac{\Delta P}{P}=2\frac{\Delta m}{M}.
\end{equation}
Since $\Delta m>0$ we see that the period of the system should
increase during the outburst if the material carries away its specific
angular momentum.

\section{The Observed Period Change in U Sco}

U Sco has a white dwarf with a mass $M_1=1.55\pm 0.24\,\rm M_\odot$
\citep{thoroughgood01}.  The radius of a non-rotating white dwarf is
given approximately by
\begin{equation}
R_1=7.99 \times 10^8 \left[ \left(\frac{M_{1}}{M_{\rm ch}}\right)^{-\frac{2}{3}}-\left(\frac{M_{1}}{M_{\rm ch}}\right)^\frac{2}{3}\right]^\frac{1}{2} \,\rm cm
\label{rad}
\end{equation}
where $M_{\rm ch}=1.44\,\rm M_\odot$ is the Chandrasekhar mass
\citep{nauenberg72}, so the radius of the white dwarf in U Sco is
$R_1=0.003\,\rm R_\odot$. We take the mass to be close to the upper
limit for that of a white dwarf that is accreting matter before a
supernova occurs, so $M_1=1.37\,\rm M_\odot$ \citep{hachisu00a}. This
mass is consistent with the fact that U Sco has such frequent
outbursts.  With equation~(\ref{dm}) we find the mass accumulated
before the outburst to be $\Delta m=2.36\times 10^{-6}\,\rm M_\odot$,
consistent with estimates by \cite{hachisu00b} and we assume that all
of this mass is ejected in the outburst. The evolved companion star
has a mass of $M_{2}=0.88 \,\rm M_\odot$ and a radius of
$R_{2}=2.1\,\rm R_\odot$ \citep{thoroughgood01}.

The orbital period of the binary before the 1999 outburst was measured
to be $P_{\rm i}=1.2305521\,\rm d$
\citep{schaefer90,schaefer95}. After the 1999 outburst the period was
measured again and was observed to be $P_{\rm f}=1.2305470\,\rm d$
(Schaefer, unpublished).  The relative change is therefore
\begin{equation}
\frac{\Delta P}{P}=\frac{P_{\rm f}-P_{\rm i}}{P_{\rm i}}=-4.1\pm 0.8 \times 10^{-6}.
\end{equation}
As we showed in the previous section, if the ejected mass carries away
its specific angular momentum then the period of the system should
increase in the outburst. We now consider mechanisms that can decrease
the orbital period, both during the outburst and between outbursts.

\subsection{Magnetic Braking}

Between outbursts, for the systems with relatively long orbital
periods, like U Sco, magnetic braking provides the largest continual
loss of angular momentum from the system. The rate of loss of angular
momentum is given roughly by
\begin{equation}
\dot J_{\rm MB}=-5.83\times 10^{-16}\left(\frac{R_1}{R_\odot}\right)^3 (\Omega \,\rm yr)^3\,\rm M_\odot R_\odot^2\,yr^{-2}
\end{equation}
\citep{rappaport83}.  The timescale on which magnetic braking operates
is
\begin{equation}
\tau_{\rm MB}=\frac{J}{\dot J_{\rm MB}},
\end{equation}
which gives for U Sco a timescale of about $2.4\times 10^9 \,\rm
yr$. The time between outbursts in U Sco may be as short as $7.9\,\rm
yr$ while the inter-eruption times for the known adjacent eruptions
are 10.8, 8.9, 10.4, 7.9, 11.8 and now $10.8\,\rm yr$ with an overall
average of $10.3\,\rm yr$ \citep{schaefer05,schaefer10b}. Because the
magnetic braking timescale is much longer, it is expected to have very
little effect on the observed orbital period. The magnetic braking
rate may about an order of magnitude lower than the rate given here
\citep{martin05}, however, this would make its timescale even
longer. Magnetic braking is more important for classical novae which
have eruptions every ten thousand years or so.

\subsection{Mass Accretion on to the Companion}

\cite{shara86} took into account the fraction of the ejected mass
$\beta$ that may be captured by the companion in the outburst and
found the separation change to be
\begin{equation}
\frac{\Delta a}{a}=\frac{\Delta m}{M_1}
\left(\frac{1+2\beta q-\beta}{1+q}-\frac{2\beta}{q(q+1)}\right),
\end{equation}
(compare to equation~\ref{over} where $\beta=0$) where
$q=M_{2}/M_1$. In the absence of strong magnetic effects the maximum
value of $\beta$ is the fractional area of the companion's accretion
radius. In order for the separation to decrease during the outburst by
mass accretion on to the companion we need
\begin{equation}
\beta<\frac{q}{2+q-2q^2}.
\end{equation}
For U Sco with $q=0.64$ this requires $\beta>0.35$ which is highly
unlikely in such a wide system.

\subsection{Frictional Angular Momentum Losses}

\cite{livio91} further considered changes to the system because of
frictional angular momentum losses as the binary moves through the
common envelope created by the ejected material. This causes the
separation of the system to decrease and so the period decreases
too. However, they found that frictional angular momentum losses are
high enough to actually cause a decrease in the separation only in the
systems with the shortest periods of around a few hours. Since U Sco
has a long orbital period of $30\,\rm hr$, it is unlikely that this
mechanism could cause the observed decrease in the orbital period.

\subsection{A Potential Alternative Explanation}

We cannot explain the observed period decrease with the usual
mechanisms for angular momentum loss either between the outbursts (by
magnetic braking) or during the outbursts (by mass accretion on to the
companion or frictional angular momentum losses). In this section we
consider whether the magnetic field of the secondary star, that is
rotating synchronously with the orbit, could provide the required
angular momentum loss during the outburst.

Suppose that the ejected mass takes away more angular momentum than
its specific angular momentum. If the ejected mass is forced to
corrotate with the binary orbit by coupling to the secondary star's
magnetic field, it would take angular momentum directly from the orbit
as it is spun up.  The angular momentum of the system before the
outburst, $J_{\rm i}$, and after the outburst, $J_{\rm f}$, are found
with equation~(\ref{angmom}) and the two observed periods. Then the
observed change in the angular momentum of the system is
\begin{equation}
  \Delta J_{\rm obs}=J_{\rm f}-J_{\rm i}.
\end{equation}
For U Sco we find that this angular momentum change is nearly four times
larger than the specific angular momentum of the ejected mass.  

The angular momentum of ejected material that corrotates with the
binary up to a radial distance $R_{\rm c}$ is given by
\begin{equation}
\Delta J= \Delta m R_{\rm c}^2\Omega_{\rm f}.
\label{jobs}
\end{equation}
Therefore, to account for the observed change in the system's angular
momentum, corrotation needs to be enforced up to
\begin{equation}
R_{\rm c}=\sqrt{\frac{(-\Delta J_{\rm obs})}{\Delta m}\frac{1}{\Omega_{\rm f}}}.
\end{equation}
For U Sco the corrotation radius is at $R_{\rm c}=5.5 {\,\rm
  R_\odot}=0.76 \, a$.  Magnetic pressure and and ram pressure of the
ejected material balance at the Alfv\'{e}n radius given by
\begin{equation}
R_{\rm A}=\left(\frac{\mu}{\dot M^2 G M_2}\right)^\frac{1}{7}
\label{ra}
\end{equation}
where $\mu$ is the dipole moment of the secondary magnetic star with
mass $M_2$ and $\dot M$ is the mass ejection rate.  The material
corrotes up to this radius and so we set $R_{\rm c}=R_{\rm A}$ and
find the required dipole moment
\begin{equation}
\mu=\left(R_{\rm c}^7 \dot M^2 G M_2\right)^\frac{1}{4}.
\end{equation}
The average mass-loss rate is
\begin{equation}
\dot M=\frac{\Delta m}{\tau}
\label{mdot}
\end{equation}
where $\tau$ is the timescale over which the mass is lost. We can take
$\tau\approx 3\,\rm months$ \citep[the timescale on which the optical
  light curve drops back to quiescence,][]{matsumoto03}. The magnetic
field strength at the stellar surface is given by
\begin{equation}
B=\frac{\mu}{R_2^3}.
\end{equation}
From equations~(\ref{ra}) to~(\ref{mdot}) we find that the secondary
star would need to have a surface field strength of $B=8.0\times
10^3\,\rm G$ in order to account for the observed period change in U
Sco.

\section{Discussion and Conclusions}

The decrease in the orbital period of U Sco during the 1999 outburst,
if confirmed, cannot be explained by evolutionary magnetic braking
between outbursts, accretion of mass on to the companion or by
frictional angular momentum losses.  However, if there is a
sufficiently strong magnetic field on the companion, then it is
possible that the ejected material may be forced to couple with the
binary orbit, thus removing angular momentum from it and decreasing
the period of the binary.

Magnetic fields of the order of a few kilogauss on the secondary star
have been suggested previously \citep[e.g.][]{meyer96,warner96}. While
such strong fields are typical of magnetic Ap stars, they may be less
common in the secondaries of cataclysmic variables. However, high
magnetic fields have been discussed for cataclysmic variables of
shorter periods by \cite{meintjes06}. The subgiant companion in U Sco
is expected to be synchronously rotating with the orbit, with a period
of $30\,\rm hr$, which is very fast for a subgiant. Studies of
late-type stars show that high fields can be expected for fast
rotators \citep[e.g. ][]{noyes84}. If the strong magnetic field of the
companion is present, the period decrease should again occur in the
2010 outburst. Measurements of the orbital period after the 2010
outburst are therefore strongly encouraged.

We should also note that orbital period changes have been observed in
binary systems not involving nova outbursts \citep[e.g. V471 Tau,
][]{skillman88}. In V471 Tau in particular, a decrease in the orbital
period of the same order of magnitude as that in U Sco has been
observed. Several authors have proposed that the decrease was caused
by a change in the internal structure of the star that changes the
non-negligible quadrupole moment \citep[e.g.][]{warner88,applegate87}.
However, as it has been shown by \cite{marsh90}, the proposed
mechanisms could only work on timescales that are longer than the
observed one (about 4 years) by more than an order of magnitude.

\cite{applegate92} also proposed a mechanism for orbital period
modulation in close, {\it non-nova}, binaries. His model relies on the
gravitational coupling of the orbit to variations in the shape of a
magnetically active star. For this model to work, however, mean
subsurface fields of several kilogauss are required. If such fields
are indeed present, then, as we have shown, the observed period change
in U Sco (which does undergo nova outbursts) can be plausibly
explained by coupling of the ejecta to the object. Limits on the
parabolic term in the O-C diagram during quiescence also seem to
indicate that Applegate's mechanism does not operate in U Sco.

\acknowledgments

We thank Jim Pringle for helpful discussions.

\clearpage

\end{document}